\documentclass[twocolumn]{aastex7}
\usepackage{amsmath}

\newcommand{\HeII}{He\,{\sc ii}}

\shorttitle{Continuum Time Delay Variation Phenomenon}
\shortauthors{Zhou et al.}

\begin{document}

\title{Unprecedented Continuum Time Delay Variation Phenomenon in an Active Supermassive Black Hole}

\correspondingauthor{Mouyuan Sun, Hai-Cheng Feng, Sha-Sha Li}
\email{msun88@xmu.edu.cn, hcfeng@ynao.ac.cn, lishasha@ynao.ac.cn}

\author[0009-0005-2801-6594]{Shuying Zhou}
\affiliation{Department of Astronomy, Xiamen University, Xiamen, Fujian 361005, People’s Republic of China}
\affiliation{SHAO-XMU Joint Center for Astrophysics, Xiamen, Fujian 361005, People’s Republic of China}
\email{zhoushuying@stu.xmu.edu.cn}

\author[0000-0002-0771-2153]{Mouyuan Sun}
\affiliation{Department of Astronomy, Xiamen University, Xiamen, Fujian 361005, People’s Republic of China}
\affiliation{SHAO-XMU Joint Center for Astrophysics, Xiamen, Fujian 361005, People’s Republic of China}
\email{msun88@xmu.edu.cn}

\author[0000-0002-1530-2680]{Hai-Cheng Feng}
\affiliation{Yunnan Observatories, Chinese Academy of Sciences, Kunming, Yunnan 650216, People’s Republic of China}
\affiliation{Key Laboratory for the Structure and Evolution of Celestial Objects, Chinese Academy of Sciences, Kunming, Yunnan 650216, People’s Republic of China}
\affiliation{Center for Astronomical Mega-Science, Chinese Academy of Sciences, Beijing 100012, People’s Republic of China}
\email{hcfeng@ynao.ac.cn}

\author[0000-0003-3823-3419]{Sha-Sha Li}
\affiliation{Yunnan Observatories, Chinese Academy of Sciences, Kunming, Yunnan 650216, People’s Republic of China}
\affiliation{Key Laboratory for the Structure and Evolution of Celestial Objects, Chinese Academy of Sciences, Kunming, Yunnan 650216, People’s Republic of China}
\affiliation{Center for Astronomical Mega-Science, Chinese Academy of Sciences, Beijing 100012, People’s Republic of China}
\email{lishasha@ynao.ac.cn}

\author[0000-0002-1935-8104]{Yongquan Xue}
\affiliation{Department of Astronomy, University of Science and Technology of China, Hefei, Anhui 230026, People’s Republic of China}
\affiliation{School of Astronomy and Space Science, University of Science and Technology of China, Hefei, Anhui 230026, People’s Republic of China}
\email{xuey@ustc.edu.cn}

\author[0000-0002-4419-6434]{Jun-Xian Wang}
\affiliation{Department of Astronomy, University of Science and Technology of China, Hefei, Anhui 230026, People’s Republic of China}
\affiliation{School of Astronomy and Space Science, University of Science and Technology of China, Hefei, Anhui 230026, People’s Republic of China}
\email{jxw@ustc.edu.cn}

\author[0000-0002-4223-2198]{Zhen-Yi Cai}
\affiliation{Department of Astronomy, University of Science and Technology of China, Hefei, Anhui 230026, People’s Republic of China}
\affiliation{School of Astronomy and Space Science, University of Science and Technology of China, Hefei, Anhui 230026, People’s Republic of China}
\email{zcai@ustc.edu.cn}

\author{Jin-Ming Bai}
\affiliation{Yunnan Observatories, Chinese Academy of Sciences, Kunming, Yunnan 650216, People’s Republic of China}
\affiliation{Key Laboratory for the Structure and Evolution of Celestial Objects, Chinese Academy of Sciences, Kunming, Yunnan 650216, People’s Republic of China}
\affiliation{Center for Astronomical Mega-Science, Chinese Academy of Sciences, Beijing 100012, People’s Republic of China}
\email{baijinming@ynao.ac.cn}

\author{Danyang Li}
\affiliation{High School Affiliated to Yunnan Normal College, Kunming, Yunnan 650506, People’s Republic of China}
\affiliation{Department of Astronomy, Xiamen University, Xiamen, Fujian 361005, People’s Republic of China}
\email{l18082932667@163.com}

\author[0000-0001-8416-7059]{Hengxiao Guo}
\affiliation{Shanghai Astronomical Observatory, Chinese Academy of Sciences, 80 Nandan Road, Shanghai 200030, People’s Republic of China}
\affiliation{SHAO-XMU Joint Center for Astrophysics, Xiamen, Fujian 361005, People’s Republic of China}
\email{hxguo@shao.ac.cn}

\author[0000-0002-2153-3688]{H. T. Liu}
\affiliation{Yunnan Observatories, Chinese Academy of Sciences, Kunming, Yunnan 650216, People’s Republic of China}
\affiliation{Key Laboratory for the Structure and Evolution of Celestial Objects, Chinese Academy of Sciences, Kunming, Yunnan 650216, People’s Republic of China}
\affiliation{Center for Astronomical Mega-Science, Chinese Academy of Sciences, Beijing 100012, People’s Republic of China}
\email{htliu@ynao.ac.cn}

\author[0000-0002-2310-0982]{Kai-Xing Lu}
\affiliation{Yunnan Observatories, Chinese Academy of Sciences, Kunming, Yunnan 650216, People’s Republic of China}
\affiliation{Key Laboratory for the Structure and Evolution of Celestial Objects, Chinese Academy of Sciences, Kunming, Yunnan 650216, People’s Republic of China}
\affiliation{Center for Astronomical Mega-Science, Chinese Academy of Sciences, Beijing 100012, People’s Republic of China}
\email{lukx@ynao.ac.cn}

\author[0000-0002-7077-7195]{Jirong Mao}
\affiliation{Yunnan Observatories, Chinese Academy of Sciences, Kunming, Yunnan 650216, People’s Republic of China}
\affiliation{Key Laboratory for the Structure and Evolution of Celestial Objects, Chinese Academy of Sciences, Kunming, Yunnan 650216, People’s Republic of China}
\affiliation{Center for Astronomical Mega-Science, Chinese Academy of Sciences, Beijing 100012, People’s Republic of China}
\email{jirongmao@mail.ynao.ac.cn}

\author[0000-0002-1380-1785]{Marcin Marculewicz}
\affiliation{Department of Astronomy, Xiamen University, Xiamen, Fujian 361005, People’s Republic of China}
\affiliation{Wayne State University, Department of Physics \& Astronomy, 666 W Hancock Street, Detroit, MI 48201, USA}
\email{marcin.marculewicz21@xmu.edu.cn}

\author[0000-0003-4156-3793]{Jian-Guo Wang}
\affiliation{Yunnan Observatories, Chinese Academy of Sciences, Kunming, Yunnan 650216, People’s Republic of China}
\affiliation{Key Laboratory for the Structure and Evolution of Celestial Objects, Chinese Academy of Sciences, Kunming, Yunnan 650216, People’s Republic of China}
\affiliation{Center for Astronomical Mega-Science, Chinese Academy of Sciences, Beijing 100012, People’s Republic of China}
\email{wangjg@ynao.ac.cn}

\begin{abstract}
Resolving the accretion disks and broad line regions (BLRs) of active galactic nuclei (AGNs) can probe the physics behind supermassive black holes (SMBHs) fueling and weigh SMBHs. With time-domain observations, the reverberation mapping (RM) technique measures time delays between different light curves, probing the AGN inner structures that are otherwise often too compact to resolve spatially with current facilities. Theoretically, the SMBH accretion disk structure does not evolve over decades. Here we report the significant variations in the continuum time delays of NGC 4151. In the high-flux state, our high-cadence ($\sim 2$ days) spectroscopy reveals that continuum time delays are $3.8^{+1.8}_{-1.0}$ times larger than those in the low-flux state and $14.9\pm 2.0$ times longer than the classical standard thin disk prediction. Notably, the continuum time delays can be comparable with the time delay between H$\beta$ and the $5100\ \mathrm{\AA}$ continuum, and the latter is commonly used to calculate the BLR sizes. Hence, the BLR sizes are underestimated if the continuum time delays are not considered, which introduces $\sim 0.3$ dex systematic uncertainties on RM SMBH masses. Our findings underscore that simultaneous continuum and BLR RMs are vital for better deciphering the SMBH accretion and mass function. 
\end{abstract}


\section{Introduction} \label{sec:intro}
Combining the linear and angular scales of the inner structures of active galactic nuclei (AGNs) allows for the measurement of parallax distances \citep{Elvis2002, Cackett2007-H0, Wang2020, GRAVITY2021} of supermassive black holes (SMBHs) at various redshift, thereby independently constrain the cosmological expansion history and resolve the Hubble tension \citep[for a review, see, e.g.,][]{hubble_tension}. The inner structures also provide valuable clues for weighing distant SMBHs \citep[e.g.,][]{GRAVITY2018} and constraining the SMBH-galaxy co-evolution \citep{Ho2013}. However, it is very challenging to resolve AGN inner structures---the central engines and broad line regions (BLRs)---directly because of their small physical sizes ($\sim$ light-day to light-month) and great distances from us. Fortunately, the Event Horizon Telescope has spatially imaged the two SMBHs, Messier 87$^*$ \citep{EHT2019} and Sagittarius A$^*$ \citep{EHT2022}, and the spectroastrometric observations of the GRAVITY instrument on board of the Very Large Telescope Interferometer (VLTI) have resolved the BLRs for several AGNs \citep[e.g.,][]{GRAVITY2018}. The upcoming GRAVITY+/VLTI will provide the angular scales of BLRs for a few hundred AGNs \citep{GRAVITY2022}. Then, it is of great importance to accurately measure the linear scales of AGN inner structures. 

AGN variability across all electromagnetic bands enables the use of the reverberation mapping technique \citep[RM;][]{Blandford1982} to probe the linear scales of AGN inner structures in the time domain. The RM technique measures the time delays between light curves of different wavelengths \citep[for a recent review, see][]{Cackett2021}, e.g., the time delays between broad emission lines (BELs) and the adjacent UV/optical continua measure the BLR linear scales (BLR RM), and the time delays between UV/optical continuum emission probe the SMBH accretion-disk sizes (continuum RM). 

BELs produced by the photoionized BLR clouds should vary in response to the extreme ultraviolet (EUV) ionizing continuum with a light travel time delay that accounts for the BLR linear scale. However, BLR RMs often measure the time delays between the BELs and the adjacent UV/optical continua and neglect the time delay between the UV/optical and EUV emission. Recent continuum RM observations \citep[e.g.,][]{Fausnaugh2016-5548, Homayouni2019-SDSS} indicate that the optical-EUV time delay is about three times larger than the expected light travel time of the classical standard thin disk \citep[SSD;][]{Shakura1973}. Then, neglecting this time delay may underestimate the BLR linear-scale measurement \citep[e.g.,][]{Pei2017-5548, Williams2020}. Quasi-simultaneous BLR and continuum RMs are required to account for this bias if the continuum time delays change from time to time. Hence, a more thorough understanding of continuum time delays and their possible variations is still warranted. 

Repeated continuum RMs for the same AGN in different flux states could offer a unique pathway for understanding the continuum time delays whose physical origin is still under active debate \citep[e.g.,][]{Chelouche2019, Cai2018, Sun2020-CHAR}. Such observations can only be performed for AGNs that exhibit dramatic flux variations. NGC 4151 is a well-known local Seyfert galaxy (redshift $z=0.00332$) with robust cosmological distance \citep[$15.8\ \mathrm{Mpc}$;][]{Yuan2020-distance} and black-hole mass measurements \citep[e.g.,][]{Onken2014-mass, Roberts2021}, and a favorable source of extragalactic neutrino \citep{Abbasi2024}. As shown by its $120$-year long-term light curve \citep{Oknyanskij2016}, NGC 4151 has large SMBH accretion-powered continuum variations accompanied by the suppression or appearance of BELs \citep[e.g.,][]{Sergeev2001, Li2022}. In addition, \citet{Edelson2017} (hereafter E17) has performed a continuum RM observation for NGC 4151 in a low-flux state ($1.04\times 10^{-14}\ \mathrm{erg\ s^{-1}\ cm^{-2}\ \textrm{\AA}^{-1}}$ at $5500\ \mathrm{\AA}$) in 2016, and now the target has entered a high-flux state since 2020 \citep[$3.42\times 10^{-14}\ \mathrm{erg\ s^{-1}\ cm^{-2}\ \textrm{\AA}^{-1}}$ at $5500\ \mathrm{\AA}$; panel (a) of Figure~\ref{fig2:light_curve}; also see][hereafter C23]{Chen2023}. Thus, NGC 4151 is an ideal target for repeated continuum RM studies. 

We report a new continuum RM campaign with spectroscopy for NGC 4151 in a high-flux state. Combined with the historical low-flux continuum RM performed by E17, we directly show that the continuum time delays vary drastically with time. The time delay observations defy the X-ray reprocessed SSD model. The significant continuum time delay variation adds substantial systematic uncertainties to the measurement of BLR linear size via the BLR RM, further affecting the black hole mass measurement and related cosmological studies. 

The manuscript is organized as follows. In Section~\ref{sec:Observations}, we present the observations and properties of NGC 4151 in the low- and high-flux states; in Section~\ref{sec:Results}, we measure the continuum RM time delays and compare the results with the low-flux state observations and BLR RM observations; in Section~\ref{sec:implication}, we show the implications of the results for AGN accretion physics, BLR sizes, black hole mass, and related cosmology; and Section~\ref{sec:summary} summaries the main conclusions.

\section{Observations and Data Reduction} \label{sec:Observations}
\subsection{Optical spectroscopic data} \label{subsec:lc_extraction}
NGC 4151 has been spectroscopically monitored over the yearly observing season, from 2022 November to 2023 July (Modified Julian Day (MJD) in $59895$--$60127$), using the Yunnan Faint Object Spectrograph and Camera mounted on the Lijiang $2.4$-$\mathrm{m}$ telescope in Yunnan Observatory, Chinese Academy of Sciences. The duration of observations is $232$ days with $88$ spectra, and the cadence ranges from $0.8$ days to $12$ days with a median cadence of $2$ days. 

The specific observation mode utilized Grism~14 with a slit width of 5\farcs05, and an additional UV-blocking filter was employed to cut off light with wavelengths below 4150 \AA, thereby reducing the impact of second-order spectra. In this observation mode, the spectral dispersion is 1.8 \AA/pixel, covering a wavelength range from 4193 \AA \ to 7374 \AA \ \citep{Feng2024}. A long slit was used, which allows for simultaneous observation of the target and a comparison star. This setup facilitates accurate flux calibration by using the comparison star.  

The raw data are processed using the standard procedures in \texttt{PyRAF},\footnote{\url{https://github.com/iraf-community/pyraf}} with an extraction aperture of 8\farcs49 and background region of 14\farcs15 to 19\farcs81. The flux calibration process involves several key steps: first, the comparison star is calibrated using observations of standard stars taken under good weather conditions to generate an initial template spectrum. This template is then matched against spectra from a stellar template library \citep{Yan2019}, selecting the spectrum with the smallest fitting residuals as the standard template. Using this standard template, along with the daily spectra of the comparison star, we perform flux calibration and atmospheric absorption correction on the target, obtaining the final flux-calibrated spectra.

\begin{figure}
    \centering
    \epsscale{1.2}
    \plotone{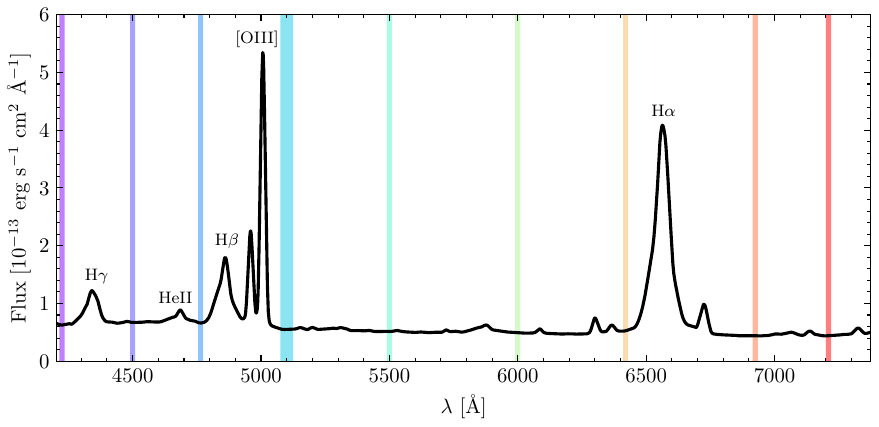}
    \caption{The rest-frame mean spectrum of NGC 4151 during the high-flux state. The colored vertical shaded areas represent the nine line-free wavelength windows listed in Table~\ref{tbl:bins}. These wavelength windows are carefully selected to avoid BEL contamination.}
    \label{fig1:continuum_window}
\end{figure}

\begin{figure*}
    \centering
    \plotone{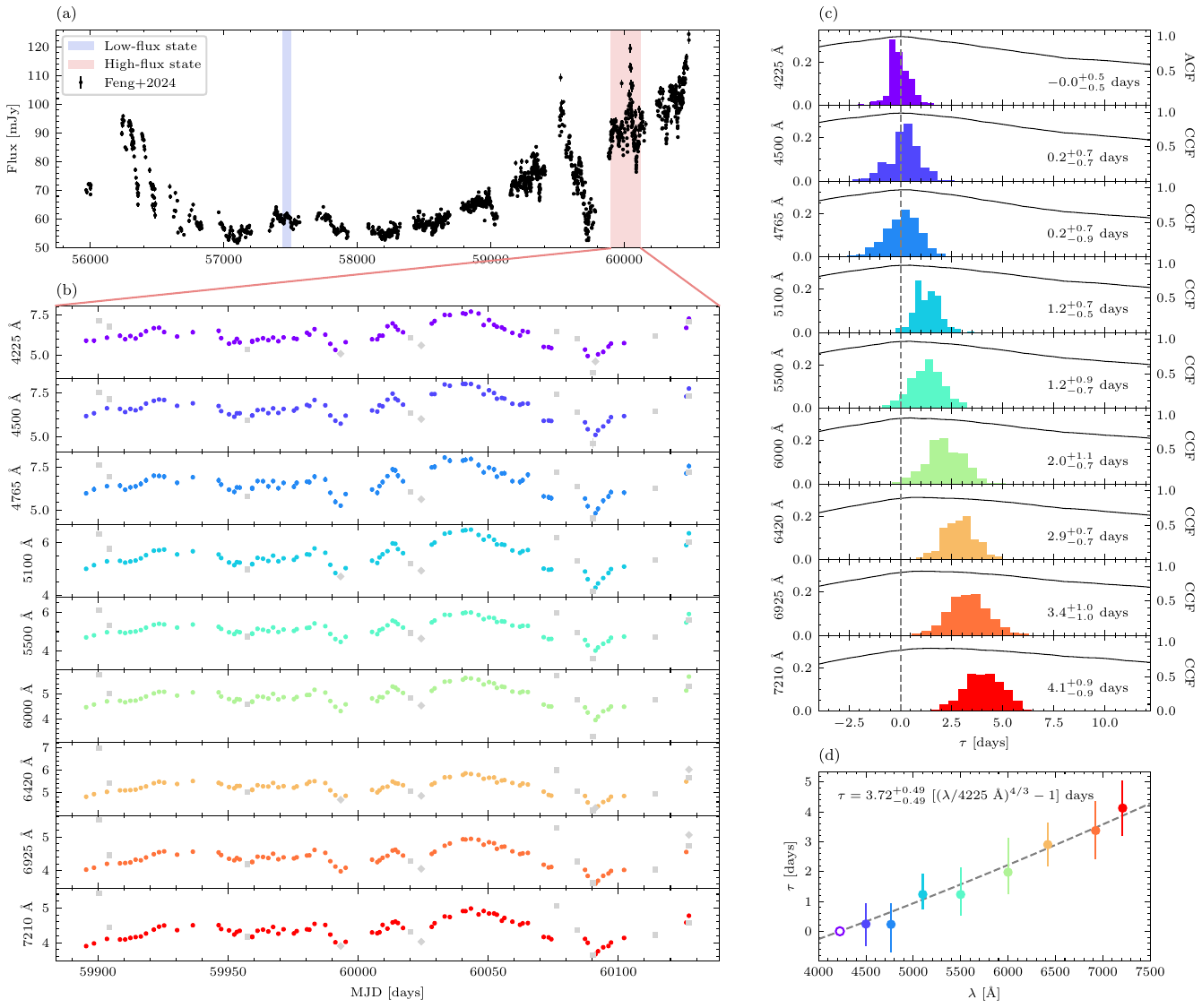}
    \caption{The multi-wavelength light curves and cross-correlation functions for NGC 4151. \textbf{Panel (a):} the long-term light curve of NGC 4151 without subtracting the host galaxy contamination. The black dots are the long-term light curve compiled by \cite{Feng2024} using \texttt{PyCALI} \citep{Li2014} from the Lijiang $2.4$-m telescope ($B$ band), ASAS-SN ($V$ and $g$ bands), and ZTF ($g$ band), where ASAS-SN $g$ band is the reference band with fluxes in units of mJy. The blue-shaded area is the period analyzed by E17 when NGC 4151 is in a low-flux state. The red-shaded area is the period analyzed in this work when NGC 4151 is in a high-flux state. \textbf{Panel (b):} the multi-band continuum light curves at the high-flux state. The fluxes are in units of $10^{-14}\ \mathrm{erg\ cm^{-2}\ s^{-1}\ \textrm{\AA}^{-1}}$. The flux error bars are too small to be visible. Gray squares are outliers determined by spectral observation logs, while gray diamonds are outliers determined by a statistical method. \textbf{Panel (c):} the auto/cross-correlation functions and centroid time-delay distributions. \textbf{Panel (d):} the time delay as a function of wavelength. Throughout this manuscript, the error bars correspond to $1\sigma$ uncertainties unless otherwise specified.} 
    \label{fig2:light_curve}
\end{figure*}

We correct for galactic extinction and the redshift of each spectrum. To extract continuum light curves, we select nine line-free continuum windows (Table~\ref{tbl:bins} and Figure~\ref{fig1:continuum_window}). The median and standard deviation of the spectral data within each window are used as the measurement values and errors, respectively, thus deriving the light curve for each bin (see Figure~\ref{fig2:light_curve}). The light curves of H$\beta$ and several other emission lines measured by the conventional method are presented by \cite{Feng2024}. To our best knowledge, NGC 4151 is the first AGN for the continuum RM study with long-term and high-cadence ground-based spectroscopy. The advantages of the continuum RM with spectroscopy are twofold. First, we can simultaneously measure continuum and BEL time delays. Second, we can properly eliminate contamination from BELs in continuum RMs. 

\begin{deluxetable}{ccc}\label{tbl:bins}
    \tablenum{1}
    \tablecaption{Rest-frame wavelengths and measured rest-frame time delays}
    \tablewidth{0pt}
    \tablehead{
    \colhead{Wavelength range} & \colhead{Central wavelength} & \colhead{Time delays}\\ \colhead{[$\textrm{\AA}$]} & \colhead{[$\textrm{\AA}$]}  & \colhead{[days]}
    } 
    \startdata
    4215--4235 & 4225 & $0.0^{+0.5}_{-0.5}$\\
    4490--4510 & 4500 & $0.2^{+0.7}_{-0.7}$\\
    4755--4775 & 4765 & $0.2^{+0.7}_{-0.9}$\\
    5075--5125 & 5100 & $1.2^{+0.7}_{-0.5}$\\
    5490--5510 & 5500 & $1.2^{+0.9}_{-0.7}$\\
    5990--6010 & 6000 & $2.0^{+1.1}_{-0.7}$\\
    6410--6430 & 6420 & $2.9^{+0.7}_{-0.7}$\\
    6915--6935 & 6925 & $3.4^{+1.0}_{-1.0}$\\
    7200--7220 & 7210 & $4.1^{+0.9}_{-0.9}$\\
    \enddata
    \tablecomments{The first column is the rest-frame wavelength ranges used to extract fluxes from the spectra, and the second column is the rest-frame central wavelengths of wavelength ranges. The third column is the rest-frame time delays relative to $4225\ \textrm{\AA}$.}
\end{deluxetable}

\subsection{Outlier rejection} \label{subsec:outlier}
Variations due to ``problematic'' observations rather than intrinsic fluctuations in the SMBH accretion disk can bias the time delay measurements. To remove the outliers, we first obtain the smoothed light curves using a smoothing function \texttt{lowess}\footnote{\url{https://www.statsmodels.org/devel/generated/statsmodels.nonparametric.smoothers_lowess.lowess}} (with the parameter $frac=0.2$), and calculate the normalized median absolute deviation $\sigma_\mathrm{lc}$ of the residuals between the smoothed and the original light curves. The statistical outliers are points that deviate from the smoothed light curve by $\geq 3\sigma_\mathrm{lc}$. To further determine the reliability of the outliers, we check the spectra and observation logs and find some problematic spectra with poor weather conditions or instrumental issues (the gray squares in Figure~\ref{fig2:light_curve}. (b)). It turns out that some statistical outliers also correspond to such problematic spectroscopic observations. The remaining statistical outliers (gray diamonds in Figure~\ref{fig2:light_curve}. (b)) have substantial deviations between photometric and spectroscopic fluxes. 

\subsection{Host galaxy contribution} \label{subsec:host}
Starlight from the host galaxy is mixed into the observed flux and should be properly subtracted. \cite{Yang2024} performed the image decomposition for NGC 4151 using the \textit{Hubble Space Telescope} (\textit{HST}) high-resolution images. Then, they obtained the host galaxy flux at $5500\ \mathrm{\AA}$ for a circular aperture of 5 arcsec (corresponding to the \textit{Swift} observations) and the $5''.05\times 8''.49$ aperture (for the spectroscopic observations in this work). The corresponding host galaxy fluxes are $2.11\times 10^{-14}\ \mathrm{erg\ s^{-1}\ cm^{-2}\ \textrm{\AA}^{-1}}$ and $1.73\times 10^{-14}\ \mathrm{erg\ s^{-1}\ cm^{-2}\ \textrm{\AA}^{-1}}$, respectively (Yang et al., private communication). 

\subsection{Equivalent width and spectral type} \label{subsec:EW}
\begin{figure}
    \centering
    \plotone{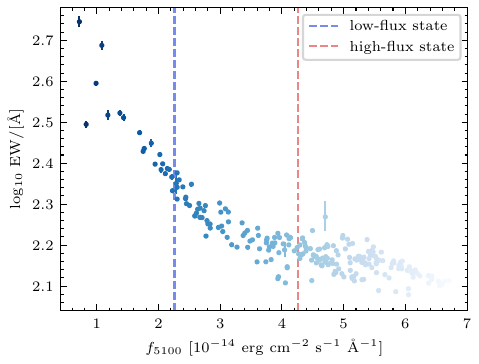}
    \caption{The evolution of EW for $\mathrm{H}\beta$ with continuum flux in NGC 4151. The x-axis is the monochromatic continuum flux for $5100\ \textrm{\AA}$. Darker colors indicate larger EWs. The vertical blue and red dashed lines represent the $5100\ \textrm{\AA}$ flux level of the low-flux state in E17 and the high-flux state in this work, respectively. }
    \label{fig3:EW}
\end{figure}

Equivalent width (EW), which is the line-to-continuum flux ratio, is an important parameter to describe the strengths of emission lines. We calculate the EWs of $\mathrm{H}\beta$ for NGC 4151 at different continuum fluxes. The definition for the EW is $\mathrm{EW}/[\textrm{\AA}]=f_\mathrm{H\beta}/f_\mathrm{5100}$, where $f_\mathrm{H\beta}$ is the integral flux for the emission line $\mathrm{H\beta}$, and $f_\mathrm{5100}$ is the monochromatic continuum flux for $5100\ \textrm{\AA}$. Data are from the four years (from $2020$ to $2024$) of the Lijiang 2.4-m telescope in Yunnan Observatory spectroscopic observations for NGC 4151 \citep[Table 2 in][]{Feng2024}. Figure~\ref{fig3:EW} shows that the H$\beta$ EW decreases with $f_{5100}$, which is consistent with the Baldwin effect \citep{Baldwin1977}. 

AGNs can be classified into different spectral types according to the line flux ratio. We follow the definition and classification of spectral types in \citet{Winkler1992}, i.e., the spectral type is decided by H$\beta$ to [O III] $5007\ \textrm{\AA}$ line ratio $f_\mathrm{H\beta}/f_\mathrm{[O\ III]}$. For the high-flux state, $f_\mathrm{H\beta}/f_\mathrm{[O\ III]}$ is $0.61$, where $f_\mathrm{H\beta}$ and $f_\mathrm{[O\ III]}$ are average emission line fluxes for the high-flux state from \cite{Feng2024}. For the low-flux state, due to the lack of simultaneous spectral observations, we use the closest spectral observations during $2018$--$2019$ monitored by the MAHA program (C23), which has a similar flux level as the low-flux state. We use the following procedures to account for the possible H$\beta$ flux offset between C23 and \cite{Feng2024}: first, we use a linear function to fit the relationship between the H$\beta$ fluxes from the two works that overlap in time; second, we use the best-fitting function to convert H$\beta$ fluxes from C23 to \cite{Feng2024}. Then, we calculate $f_\mathrm{H\beta}/f_\mathrm{[O\ III]}=0.28$ for the low-flux state, where $f_\mathrm{H\beta}$ is the average value of the calibrated C23 H$\beta$ flux, and $f_\mathrm{[O\ III]}$ is the same as the high-flux state. Thus, the changing-look AGN (CLAGN) NGC 4151 changes from the low-flux Type 1.8 state to the high-flux Type 1.5 state.

\subsection{Swift/XRT observations}
\label{subsec:swift-xrt}
NGC 4151 was frequently observed by the X-Ray Telescope (XRT) on board \textit{Swift}. To obtain the X-ray light curve that covers the low- and high-flux states of NGC 4151, we consider the \textit{Swift}/XRT observations with the target IDs $00034455$ and $00096883$. The XRT light curve is built via the UK Swift Science Data Centre online tool \citep{Evans2007-XRT}. We calculate the median XRT count rates for the MJD ranges $[57438, 57507]$ (the low-flux state) and $[59890, 60100]$ (the high-flux state), respectively. The $0.3$-$10$ keV median count rates for the two states are $1.03\pm 0.02\ \mathrm{counts\ s^{-1}}$ and $1.7\pm 0.1\ \mathrm{counts\ s^{-1}}$, respectively. Hence, the X-ray flux varies by a factor of $1.65\pm 0.07$, which can cause the disk size to increase by a factor of $1.65^{1/3}=1.18$ for the X-ray reprocessing of an SSD (see Section~\ref{subsubsec:compare_low}). 

NGC 4151 is also included in the \textit{Swift}/BAT $70$-month AGN Catalog with the intrinsic $14$--$195$ keV luminosity of $10^{43.1}\ \mathrm{erg\ s^{-1}}$ \citep{Ricci2017-BASS}. The ratio of the X-ray to disk luminosities is $\simeq 20\%$. Therefore, X-ray illumination should not be the main energy source for disk emission.

\subsection{Target properties} \label{subsec:properties}
We can now summarize the luminosity and black-hole mass of NGC 4151. For the high-flux state in this work (the red-shaded area in Panel (a) of Figure~\ref{fig2:light_curve}), we take the average of the fluxes in rest-frame $5490$--$5510\ \textrm{\AA}$ band as the flux $f_{5500}$ at $5500\ \textrm{\AA}$, which has a value of $3.42\times 10^{-14}\ \mathrm{erg\ s^{-1}\ cm^{-2}\ \textrm{\AA}^{-1}}$ after subtracting the host galaxy contamination (Section~\ref{subsec:host}). The bolometric luminosity $L_\mathrm{bol}$ can be estimated using $f_{5500}$, combined with the distance measured by Cepheid stars \citep[$15.8\ \mathrm{Mpc}$;][]{Yuan2020-distance} and a bolometric correction factor of $10$ \citep{Richards2006-BC}. Thus, the bolometric luminosity $L_\mathrm{bol}$ for the high-flux state is $5.61\times 10^{43}\ \mathrm{erg\ s^{-1}}$. E17 used \textit{Swift} UVOT to monitor NGC 4151 in a low-flux state for UV/optical bands (rest-frame $1645$--$5781\ \textrm{\AA}$; the blue line range in Panel (a) of Figure~\ref{fig4:tau_evolve}) over 69 days (MJD in $57507$--$57438$; the blue-shaded area in Panel (a) of Figure~\ref{fig2:light_curve}) with high-frequency sampling $\sim 0.2\ \mathrm{days}$. The flux $f_{5500}$ for the low-flux state is $1.04\times 10^{-14}\ \mathrm{erg\ s^{-1}\ cm^{-2}\ \textrm{\AA}^{-1}}$, estimated from the average flux in the $V$ band (observed-frame $5050$--$5800\ \textrm{\AA}$) after subtracting the galaxy contribution. The bolometric luminosity for the low-flux state is $1.71\times 10^{43}\ \mathrm{erg\ s^{-1}}$. We did not report the measurement uncertainties of $L_{\mathrm{bol}}$ because the systematic uncertainties \citep[about a factor of two; see, e.g.,][]{Richards2006-BC} dominate the error budget. The optical-based bolometric luminosity is also consistent with the hard X-ray observations (Section~\ref{subsec:swift-xrt}). The luminosity ratio $\mathcal{R}_L$ between the high- and low-flux states is $3.27$. The Eddington ratios $\dot{m}=L_\mathrm{bol}/L_\mathrm{Edd}$ are $0.003$ and $0.01$ for the low- and high-flux states, respectively, where $L_\mathrm{Edd}=1.26\times 10^{38}M_\mathrm{BH}/M_\odot\ \mathrm{erg\ s^{-1}}$ is the Eddington luminosity, and $M_\mathrm{BH}=4.27(\pm1.31)\times 10^7\ M_\odot$ is the black hole mass \citep{Onken2014-mass, Roberts2021}. NGC 4151 is an ideal target for probing the possible evolution of the accretion disk structure.

\section{Analysis} \label{sec:Results}
\subsection{Continuum RM analysis} \label{subsec:obs_analysis}
\subsubsection{Time delay measurements} \label{subsubsec:delay_measurement}
The continuum RM technique measures the time delays between different continuum wavelengths, which probe the AGN central engine structure. The time delays between observed light curves of different wavelengths can be derived using the interpolation cross-correlation function \citep[CCF;][]{Peterson1998}, which linearly interpolates the observed light curves and calculates the correlation coefficient $r$ between the light curves at different time delay shifts. We estimate time delays using \texttt{PYCCF} \citep{Sun2018_pyccf}, a widely used Python code based on the interpolated CCF. The time delays are calculated from the centroid of the interpolated CCF as the $r$-weighted delay for $r>0.8\ r_\mathrm{max}$, where $r_\mathrm{max}$ is the maximum correlation coefficient. The time delay range for calculating the interpolated CCF is $-20$ days to $20$ days, and the step is $0.5$ days. Following the flux randomization and random subset selection procedures of \cite{Cackett2018-4593}, we perform 500 Monte Carlo iterations to obtain the cross-correlation centroid distributions (CCCDs). The $50\%$, $15.87\%$, and $84.13\%$ percentiles of the CCCDs are set as the measured time delay, $1\sigma$ lower and upper limits, respectively. Panel (c) of Figure~\ref{fig2:light_curve} and Table~\ref{tbl:bins} show the rest-frame time delays for different wavelengths relative to the minimum rest-frame wavelength ($4225\ \mathrm{\AA}$) in this work. There is no significant long-term variability during the campaign. Consequently, the time-delay measurements change only slightly (by $\sim 0.1$ days) if the light curves are detrended using a first-order polynomial function. The BEL-free time delays increase with wavelengths. 

We aim to quantitatively study the relationship between time delay ($\tau$, which is with respect to a given reference wavelength $\lambda_0$) and wavelength ($\lambda$). The fitting equation is 
\begin{equation}\label{equ:tau-wave fitting}
    \tau = \tau_0(\lambda_0) [(\lambda/\lambda_0)^{\beta}-1],
\end{equation}
where $\lambda_0$ is the reference wavelength (i.e., $4225\ \textrm{\AA}$), $\tau_0(\lambda_0)$ and $\beta$ are free parameters. The AGN SSD model has an effective temperature profile $T_\mathrm{eff}(R)\propto M_\mathrm{BH}^{1/4} \dot{M}^{1/4} R^{-3/4}$, where $\dot{M}$ is the mass accretion rate. For a given wavelength $\lambda$, the corresponding characteristic radius $R(\lambda)$ is determined by $hc/\lambda = kT_\mathrm{eff}(R(\lambda))$, where $h$ is the Planck constant, $c$ is the speed of light and $k$ is Boltzmann constant, respectively. Then, the characteristic radius $R(\lambda)\propto M_\mathrm{BH}^{1/3} \dot{M}^{1/3} \lambda^{4/3}$. Hence, we fix $\beta$ to $4/3$, which is a common practice in AGN continuum RM studies (e.g., E17).

We use the maximum likelihood method and the Markov Chain Monte Carlo (MCMC) chain to obtain the best-fitting relationship and parameter uncertainties. The logarithmic likelihood function is
\begin{equation} \label{equ:likelihood function}
    \ln \mathcal{L}= -0.5\sum \left [\frac{(f_\mathrm{n}-f_\mathrm{model,n})^2}{\sigma_\mathrm{n}^2}+\ln{\sigma_\mathrm{n}^2}\right],
\end{equation}
where $f_\mathrm{n}$ is the measured time delay, and $f_\mathrm{model,n}$ is the model value. The total variance $\sigma_\mathrm{n}^2=\sigma_\mathrm{n,err}^2+(\sigma_\mathrm{int}f_\mathrm{model,n})^2$, where $\sigma_\mathrm{n,err}$ is the uncertainty for $f_\mathrm{n}$, and $\sigma_\mathrm{int}f_\mathrm{model,n}$ represent the intrinsic scatter. We set uniform prior distributions for the parameter $\tau_0(\lambda_0)$, using the MCMC code \texttt{emcee} \citep{emcee-Foreman2013} and the likelihood function to obtain the posterior distribution of the parameter. The best-fitting value for the parameter is the median of the posterior distribution, and $1\sigma$ uncertainties are taken as $16\mathrm{th}$--$84\mathrm{th}$ percentiles of the posterior distribution.
Panel (d) of Figure~\ref{fig2:light_curve} shows the fitting result for Eq.~\ref{equ:tau-wave fitting}, with the best fitting $\tau_0(4225\ \textrm{\AA})$ and its $1\sigma$ uncertainty being $3.72\ \mathrm{days}$ and $0.49\ \mathrm{days}$, respectively. The reduced $\chi^2$ of the best fit is $0.1$. Thus, same as the low-flux state analyzed in E17, Eq.~\ref{equ:tau-wave fitting} with $\beta=4/3$ can fit the relationship between time delays and wavelengths well. 

\subsubsection{Comparison with the low-flux state} \label{subsubsec:compare_low}
\begin{figure*}
    \centering
    \plotone{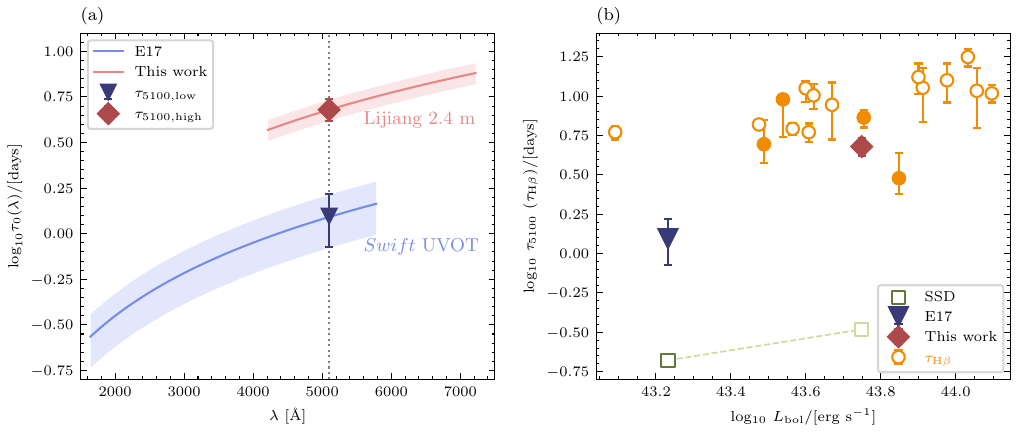}
    \caption{Time delays as a function of wavelength and luminosity. \textbf{Panel (a):} The rest-frame absolute delay versus wavelength (i.e., $\tau_0(\lambda)=\tau_0(\lambda_0)(\lambda/\lambda_0)^{4/3}$) for E17 (blue curve) and this work (red curve). The shaded areas correspond to the $1\sigma$ uncertainties. The vertical dotted line indicates $\lambda=5100\ \textrm{\AA}$. The dark-purple triangle and the red diamond are the low-flux $\tau_0(5100\ \textrm{\AA})$ ($\tau_\mathrm{5100,low}$) and high-flux states $\tau_0(5100\ \textrm{\AA})$ ($\tau_\mathrm{5100,high}$), respectively. It is evident that $\tau_\mathrm{5100,high}$ is larger than $\tau_\mathrm{5100,low}$ by a factor of $3.8^{+1.8}_{-1.0}$. \textbf{Panel (b):} The evolution of $\tau_{5100}$ (i.e., $\tau_\mathrm{disk}$ for $5100\ \textrm{\AA}$) and $\tau_\mathrm{H\beta}$ (i.e., $\tau_\mathrm{disk-BLR}$ between $5100\ \textrm{\AA}$ and H$\beta$) with the bolometric luminosity $L_\mathrm{bol}$. The purple triangle represents $\tau_\mathrm{5100,low}$, and the red diamond indicates the high-flux state $\tau_\mathrm{5100,high}$ obtained by this work. The dark- and light-green squares correspond to the theoretical $\tau_{5100}$ according to the X-ray reprocessed SSD for low- and high-flux states, respectively. The open-yellow dots are historical $\tau_\mathrm{H\beta}$ measurements (collected by C23), and the filled-yellow dots show the $\tau_\mathrm{H\beta}$ results from \cite{Feng2024}. It is clear that $\tau_{5100}$ can sometimes be comparable to $\tau_\mathrm{H\beta}$.}
    \label{fig4:tau_evolve}
\end{figure*}

NGC 4151 has continuum RM measurements for both low-flux state (E17) and high-flux state (this work). Panel (a) in Figure~\ref{fig4:tau_evolve} shows the comparison of the best-fitting absolute delay (i.e., $\tau(\lambda)=\tau_0(\lambda_0)(\lambda/\lambda_0)^{4/3}$) for the low- and high-flux states. For the low-flux state, E17 obtained the rest-frame $\tau_0(\lambda_0)=(0.34\pm 0.11)\ \mathrm{days}$ for $\lambda_0=1922\ \textrm{\AA}$. For the high-flux state, the rest-frame $\tau_0(\lambda_0)=(3.72\pm 0.49)\ \mathrm{days}$ for $\lambda_0=4225\ \textrm{\AA}$ according to Figure~\ref{fig2:light_curve}. The continuum RM measurements in low- and high-flux states differ significantly. Mrk 110 \citep{Vincentelli2022}, Mrk 335 \citep{Kara2023}, and Fairall 9 \citep{Edelson2024} also have repeated continuum RMs, and no obvious time-delay variation is detected; the UV/optical delays in Mrk 817 vary by a factor of two in adjacent epochs \citep{Lewin2024}. 

For the convenience of subsequent discussions, we calculate $\tau_0(5100\ \textrm{\AA})=\tau_0(\lambda_0)(5100/\lambda_0)^{4/3}$ and donate as $\tau_{5100}$, i.e., the rest-frame time delay between the rest-frame $5100\ \textrm{\AA}$ and a sufficiently small wavelength (i.e., $\lambda \sim 10\ \mathrm{\AA}$). Panel (b) in Figure~\ref{fig4:tau_evolve} compares $\tau_{5100}$ for the low-flux state ($\tau_\mathrm{5100,low}=1.25\pm0.40\ \mathrm{days}$; purple triangle) and the high-flux state ($\tau_\mathrm{5100,high}=4.78\pm0.63\ \mathrm{days}$; red diamond). 

The measured time delays are often compared with the X-ray reprocessed SSD model. In this model, the inter-wavelength time delays are explained as the light travel time differences from the X-ray corona to the disk UV/optical emission regions \citep[e.g.,][]{Krolik1991}. It is often assumed that the ratio of the external heating from the X-ray illumination to the disk internal heating at each radius is $k_\mathrm{X}=1/3$ \citep{Fausnaugh2016-5548}. Note that the X-ray flux as measured by the \textit{Swift} XRT increases only by a factor of $1.65$ from the low- to high-flux states (see Section~\ref{subsec:swift-xrt}). It is reasonable to assume that the mean value of $k_\mathrm{X}$ in the low-flux state is similar to that in the high-flux state. For the X-ray reprocessed SSD model, the effective temperature profile $T_\mathrm{eff}$ \citep[e.g.,][]{Fausnaugh2016-5548} is 
\begin{equation} \label{equ:T_eff}
    T_\mathrm{eff}(R) = \left (\frac{3(1+k_\mathrm{X})GM_\mathrm{BH}\dot{M}}{8\pi\sigma R^3}\right )^{1/4},
\end{equation}
where $G$, $\dot{M}=10\dot{m}L_\mathrm{Edd}/c^2$, $\sigma$, and $R$ are the Gravitational constant, accretion rate, the Stefan-Boltzmann constant, and the disk radius, respectively. The radiation characteristic radius $R(\lambda)$ for a given wavelength $\lambda$ can be calculated based on $hc/\lambda = kT_\mathrm{eff}(R(\lambda))$,
\begin{equation} \label{equ:R(lambda)}
    R(\lambda)=\left (\frac{k\lambda}{hc}\right)^{4/3} \left (\frac{3(1+k_\mathrm{X})GM_\mathrm{BH}\dot{M}}{8\pi\sigma}\right )^{1/3}.
\end{equation}
The time delay measured by the continuum RM is the time delay of the average radius weighted by fluxes, which is $\tau(\lambda)=2.49^{4/3}R(\lambda)/c$ \citep{Fausnaugh2016-5548}. Thus, the theoretical $\tau_{5100}$ based on the X-ray reprocessed SSD model (squared dots in Figure~\ref{fig4:tau_evolve}) are $0.21$ and $0.33$ days for the low- and high-flux states, respectively. The calculations are insensitive to the assumed $k_X$. The two estimates of SSD time delays are uncertain by a factor of $(2\times 2)^{1/3}\simeq 1.58$ if $M_{\mathrm{BH}}$ and $\dot{M}$ are uncertain by a factor of two. Note that the uncertainty of $M_{\mathrm{BH}}$ is possibly less than the assumed one because it is consistently measured by stellar dynamics \citep[e.g.,][]{Onken2014-mass} and BLR modeling \citep{Bentz2022}. While $\tau_\mathrm{5100,low}$ is already $6.0\pm 1.9$ times the prediction (dark-green square) of the X-ray reprocessed SSD model, $\tau_\mathrm{5100,high}$ is $14.9\pm 2.0$ times the prediction (light-green square). Hence, NGC 4151 shows the largest discrepancy among Seyfert AGNs between the measured delays and the SSD predictions. We discuss the possible reasons for the significant apparent time-delay variations in Section~\ref{subsec:variability}.

\subsection{Comparison with the BLR RM} \label{subsec:compare_BLR}
Optical emission (such as the $5100\ \textrm{\AA}$ continuum) is often used as the proxy for the ionizing continuum (such as the EUV emission) to measure the BLR locations. That is, BLR RM observations often ignore the time delay between the ionizing continuum and the optical emission; according to the SSD model, this delay is supposed to be relatively small compared with the time delays between the optical emission and BELs. However, recent observations \citep[e.g.,][]{Fausnaugh2016-5548, Homayouni2019-SDSS} have shown that $\tau_{5100}$ is about three times larger than the SSD prediction, indicating that the time delay between the ionizing continuum and the optical emission may not be negligible. NGC 4151 has many BLR RMs \citep{Maoz1991, Kaspi1996, Bentz2006, De_Rosa2018, Li2022, Chen2023, Feng2024}. The yellow dots in Panel (b) of Figure~\ref{fig4:tau_evolve} show the H$\beta$ time delays collected by C23 and four new measurements from Table 2 of \cite{Feng2024}. Surprisingly, $\tau_{5100}$ can sometimes be comparable with $\tau_{\mathrm{H\beta}}$---the rest-frame time delay between the rest-frame $5100\ \textrm{\AA}$ and $\mathrm{H\beta}$.

\begin{figure}
    \centering
    \plotone{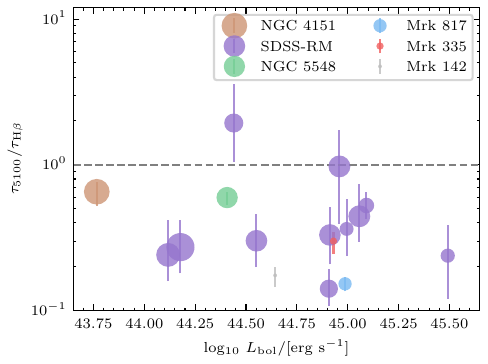}
    \caption{The ratios between $\tau_{5100}$ and $\tau_\mathrm{H\beta}$ for sources with simultaneous continuum and BLR RMs as a function of the logarithmic bolometric luminosity. The purple dots are the targets from the SDSS-RM project. The brown, green, blue, red, and gray dots are NGC 4151, NGC 5548, Mrk 817, Mrk 335, and Mrk 142, respectively. The size of the dot decreases with increasing the Eddington ratio. To obtain accurate BLR sizes, one cannot ignore $\tau_{5100}$ because $\tau_{5100}$ can be comparable to or even larger than $\tau_\mathrm{H\beta}$. The ratio $\tau_{5100}/\tau_\mathrm{H\beta}$ exhibits substantial dispersion for different targets.}
    \label{fig5:disk_vs_blr}
\end{figure}

We collect sources with simultaneous continuum and BLR RMs \citep{Bentz2015-817, Shen2015-SDSS, Fausnaugh2016-5548, Grier2017-SDSS, Pei2017-5548, Homayouni2019-SDSS, Cackett2020-142, Kara2021-817, Kara2023, Khatu2023-142, Li2024-SARM} and find that $\tau_{5100}$ can be a substantial fraction of $\tau_\mathrm{H\beta}$ (Figure~\ref{fig5:disk_vs_blr}) for some AGNs. The ratio $\tau_{5100}/\tau_\mathrm{H\beta}$ exhibits a large dispersion for different targets. This leads to substantial underestimation for BLR sizes (for further discussion, see Section~\ref{subsec:BLR_loc}), and this underestimation cannot be easily corrected. 

\section{Discussion} \label{sec:implication}
\subsection{Possible origins of the time-delay variations} \label{subsec:variability}
The SMBH accretion disk-size excess has been widely observed in continuum RM \citep[e.g.,][]{Fausnaugh2016-5548, Cackett2018-4593, Montano2022-4395, Guo2022} and quasar microlensing studies \citep[e.g.,][]{Morgan2010}. The measured delays in continuum RMs are $2$--$4$ times larger than the theoretical time delays of the X-ray reprocessed SSD model. Some models are proposed to account for the disk-size excess. The most popular model is the SDD+DC model, which assumes that the observed variable continua arise from a superposition of emission from the disk and the diffuse continuum (DC) and other components (e.g., emission lines) at the BLR scale. Indeed, the observed time delays often show the $u$-band excess resembling the Balmer jump of the DC emission \citep[e.g.,][]{Fausnaugh2016-5548, Cackett2018-4593, Lawther2018-BLR, Chelouche2019, Korista2019-BLR, Netzer2022}. The influence due to BLR DC might increase with the EW of the broad H$\beta$ \citep[e.g.,][]{Li2021-CHAR}, with the latter being essentially the line-to-continuum flux ratio. Observations show that the EW of H$\beta$ of NGC 4151 decreases with the continuum flux (Section~\ref{subsec:EW}), consistent with the Baldwin effect \citep{Baldwin1977}. However, it should be pointed out that the BLR DC originates from regions significantly smaller than those producing H$\beta$ \citep[e.g.,][]{Netzer2020, Guo2022}. Then, a lack of correlation between the BLR DC contribution and the EW of H$\beta$ cannot be excluded. Currently, robust indicators to quantify the BLR DC fraction remain elusive. Nevertheless, the observed time-delay variations can be easily explained if the DC fraction increases significantly from the low-flux to high-flux states. 

As revealed by frequency-resolved delays \citep{Cackett2022, Lewin2024, Chen2024-CHAR}, the duration and cadence (along with signal-to-noise ratio $S$/$N$ and wavelengths of the observations) can also induce apparent time-delay variations. The light curve duration and cadence of this work are longer than those of E17, which may cause the light curves in this work to contain more long-term variations compared to E17. These differences may account for the observed time-delay variations between this work and E17. Meanwhile, the observed wavelengths in this work are longer than E17, and the long-wavelength emission inherently contains more low-frequency variability than the short-wavelength emission \citep[e.g.,][]{MacLeod2010}. The wavelength differences may partially compensate for the time-delay measurement variations caused by duration and cadence. The net effect requires quantitative assessment via Monte Carlo simulations incorporating disk/BLR structures and their relative contributions. However, the DC fraction and BLR structures cannot be accurately determined. Hence, we cannot rule out the possibility that the observed time-delay variations are (at least partially) caused by the impacts of observing conditions. 

The third possibility to explain the time-delay variations involves disk structure changes between flux states. The SSD model assumes blackbody radiation and an effective temperature profile $T_\mathrm{eff}\propto R^{-3/4}$ (Eq.~\ref{equ:T_eff}). Hence, one solution is to enlarge the disk sizes by flattening the SSD temperature profile with some mechanisms, increasing the light travel time delays of the emission regions for a given wavelength. Some candidate mechanisms have been proposed, e.g., the inhomogeneous disk \citep{Dexter2011}, the windy disk \citep{Sun2019, Li2019}, and the disk embedded with stellar black holes \citep{Zhou2024}. These models still rely on X-ray reprocessing to generate inter-band time delays. The X-ray reprocessing scenario has some weaknesses. First, there is no significant correlation between the observed X-ray and UV/optical light curves \citep[e.g.,][]{Edelson2019, Cackett2021}. Second, this mechanism has an energy budget problem for the X-ray emission being too weak to drive UV/optical variations \citep{Clavel1992, Dexter2019, Marculewicz2023, Secunda2024}. Alternative AGN variability mechanisms, e.g., the inhomogeneous disk with a speculated common large-scale temperature fluctuation \citep{Cai2018} or the corona-heated accretion-disk reprocessing model \citep[CHAR;][]{Sun2020-CHAR, Chen2024-CHAR}, are proposed to account for continuum time delays. Whether these models can generate flux-state-dependent temperature profiles sufficient to explain the observed delay variations (Figure~\ref{fig4:tau_evolve}) remains unclear. 

NGC 4151 is a CLAGN. The physical mechanism for changing-look behavior in CLAGNs, i.e., the appearance or disappearance of BELs over months to years, remains unclear. CLAGNs often show evident mid-infrared variations \citep[e.g.,][]{Sheng2017}, which strongly suggests that simple obscurations cannot explain the changing-look phenomenon but are instead due to the intrinsic luminosity changes. If the observed time-delay variations are indeed caused by accretion disk or BLR variations, they could provide critical tests for various CLAGN models \citep[e.g.,][]{Ross2018, Sniegowska2020, Dexter2019b-magnetic, Feng2021}.

\subsection{BLR linear sizes} \label{subsec:BLR_loc}
\begin{figure*}
    \centering
    \plotone{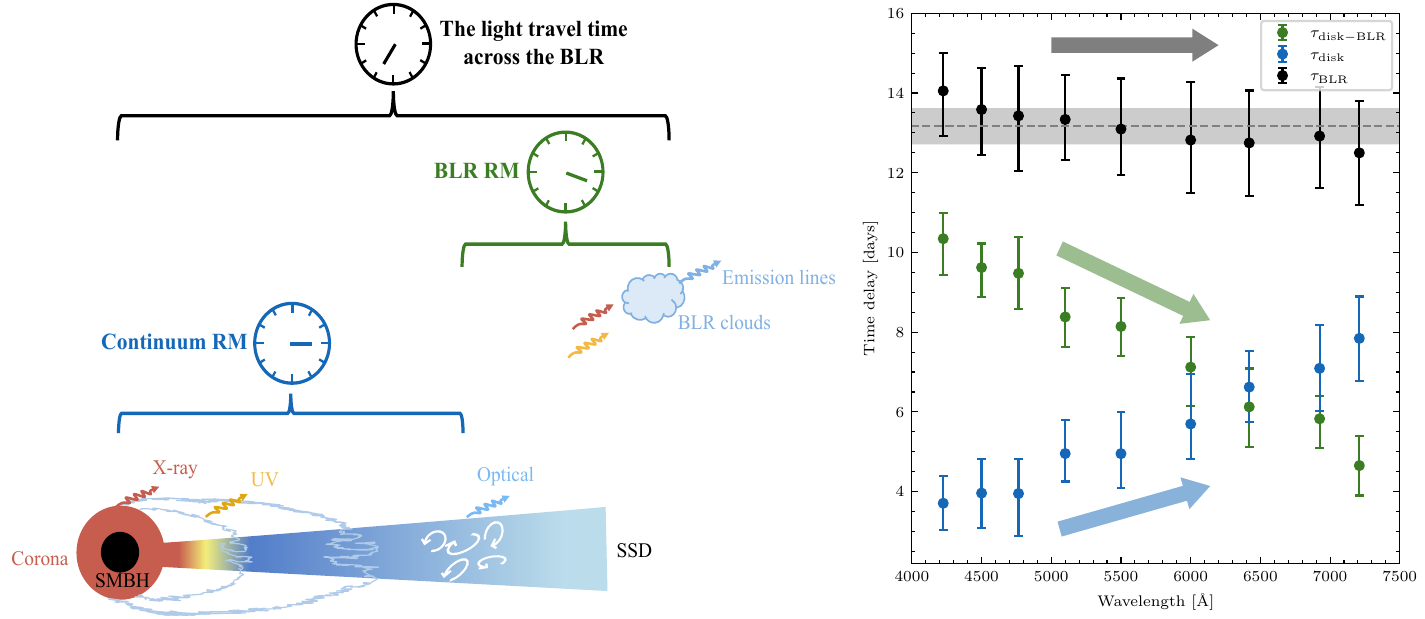}
    \caption{The BLR RM can significantly underestimate the BLR linear size if continuum time delays are neglected. Left: A schematic diagram of the continuum and BLR RMs. Right: While the continuum time delays ($\tau_\mathrm{disk}$; blue dots) increase with wavelength, the broad $\mathrm{H\beta}$ time delays ($\tau_\mathrm{disk-BLR}$; green dots) exhibit the opposite tendency. The sums ($\tau_\mathrm{BLR}$; black dots) of the continuum and $\mathrm{H\beta}$ time delays, which probe the true BLR linear scale, are insensitive to wavelength.} 
    \label{fig6:schematic}
\end{figure*}

In BLR RMs, optical emission is used as a proxy for the ionizing continuum. As a result, the time delay between the BEL and the ionizing continuum ($\tau_{\mathrm{BLR}}=R_{\mathrm{BLR}}/c$), which probes the real BLR size ($R_{\mathrm{BLR}}$), should be 
\begin{equation} \label{equ:tau_BLR}
    \tau_\mathrm{BLR}=\tau_\mathrm{disk}+\tau_\mathrm{disk-BLR}=\tau_\mathrm{disk-BLR}\left( \frac{\tau_\mathrm{disk}}{\tau_\mathrm{disk-BLR}}+1\right)\\,
\end{equation}
where $\tau_\mathrm{disk}$ and $\tau_\mathrm{disk-BLR}$ are the continuum time delay between ionizing continuum and optical emission and the time delay between the optical emission and the BEL, respectively. The two time delays can be measured via the continuum and BLR RMs (Figure~\ref{fig6:schematic}), respectively. As shown in Figures~\ref{fig4:tau_evolve} and \ref{fig5:disk_vs_blr},  $\tau_\mathrm{disk}$ can be comparable to $\tau_\mathrm{disk-BLR}$ in NGC 4151, NGC 5548, and some SDSS-RM quasars, causing the BLR linear scales to be occasionally underestimated by $\lesssim 0.3$ dex. Indeed, the time delays ($\tau_\mathrm{disk-BLR}$) between the emission line $\mathrm{H\beta}$ and various continuum emission increase with decreasing wavelengths, and the sums of the continuum and $\mathrm{H\beta}$ time delays, which correspond to $\tau_\mathrm{BLR}$ (Eq.~\ref{equ:tau_BLR}), are $13.2$ days (right panel in Figure~\ref{fig6:schematic}), i.e., substantially larger than $\tau_\mathrm{disk-BLR}$. The underestimation of the BLR size in our RM monitoring should be larger than in E17. Hence, the continuum time delay variations may contribute substantially to the scatter in the AGN $R_{\mathrm{BLR}}$--luminosity relation \citep{Bentz2013, Du2018}, and the latter is fundamental to the widely-used single-epoch virial SMBH mass estimators. 

\subsection{Black hole mass measurement} \label{subsec:BH_mass}
The location and velocity of BLR clouds can probe the central SMBH mass. If the BLR clouds are virialized, the virial mass of the central SMBH is
\begin{equation}
    M_\mathrm{BH}=fM_{\mathrm{VP}}=f\frac{c\tau_\mathrm{BLR}(\Delta V)^2}{G}\\,
\end{equation}
where $f$ is a dimensionless virial factor that includes the unknown geometries and kinematics of the BLR clouds, $M_{\mathrm{VP}}$ is the virial product (which is a measurable quantity in BLR RM studies), and $\Delta V$ is the BLR cloud velocity as represented by the line widths of the broad emission lines. As mentioned in Section~\ref{subsec:BLR_loc}, BLR RMs use $\tau_\mathrm{opt-BLR}$ to represent $\tau_\mathrm{BLR}$. The measured $M_\mathrm{VP}$ will be underestimated by \citep[see also][]{Pei2017-5548, Williams2020} a factor of ($\tau_\mathrm{disk}/\tau_\mathrm{disk-BLR}+1$). Our results demonstrate that the underestimation of $M_\mathrm{VP}$ is time dependent since $\tau_\mathrm{disk}/\tau_\mathrm{disk-BLR}$ varies significantly for the same target (Figure~\ref{fig4:tau_evolve}). This effect may be able to account for the large variations in the virial products of NGC 4151 \citep[e.g.,][]{Li2022, Chen2023}.  

The SMBH mass is $M_\mathrm{BH}=fM_{\mathrm{VP}}$, and $f$ is often calibrated according to independent SMBH mass measurements for local AGNs. Observationally speaking, it is a common approach to assume that all AGNs have similar $f$. However, our result suggests that the inferred $f$ for the same AGN is likely to be variable up to a factor of two simply because of the continuum time delay variation. Hence, the continuum time delays may contribute significantly to the observed scatter in $f$ and the systematic uncertainties in $M_{\mathrm{BH}}$. 

\begin{figure}
    \centering
    \plotone{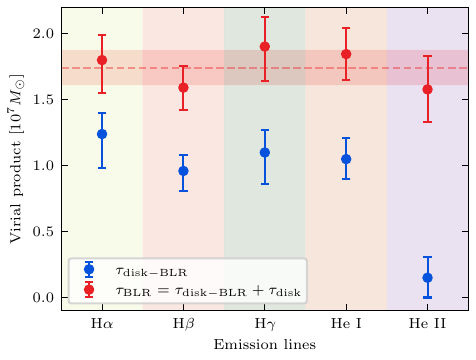}
    \caption{The continuum time delays affect the RM virial product measurements in NGC 4151. The BLR size and the corresponding virial product (the SMBH mass divided by a constant; blue dots) can be significantly underestimated because the continuum time delays are neglected. The underestimation is large for high-ionization lines, e.g., \HeII. When the continuum time delays are properly accounted for, the virial products of all lines are statistically consistent (red dots). The dashed line and shaded area represent the average value of the red dots and its $1\sigma$ uncertainty.} 
    \label{fig7:mass}
\end{figure}

The underestimation of $\tau_{\mathrm{BLR}}$ and $M_{\mathrm{VP}}$ can be larger in high-ionization BELs (e.g., \HeII$\,\lambda 4686$) than in low-ionization ones because the emission regions of the former are closer to the SMBH than the latter. Consequently, the \HeII-based $M_{\mathrm{VP}}$ is smaller than the H$\beta$-based one. For NGC 4151, the time delays of broad $\mathrm{H \alpha}$, $\mathrm{H \beta}$, $\mathrm{H\gamma}$, He\,{\sc i} $\lambda5876$, and \HeII\ with respect to the $g$-band are obtained for NGC 4151 using observations from Nov. 2022 to June 2023 \citep[Table 4 in][]{Feng2024}, which are $10.6^{+1.3}_{-2.2}$ days, $7.3^{+0.9}_{-1.1}$ days, $6.5^{+1.0}_{-1.4}$ days, $6.38^{+0.91}_{-0.88}$ days, and $0.49^{+0.52}_{-0.51}$ days, respectively; the velocity dispersions of the r.m.s. spectra for $\mathrm{H \alpha}$, $\mathrm{H \beta}$, $\mathrm{H\gamma}$, He\,{\sc i}, and  \HeII\ are $2445\pm 48\ \mathrm{km\ s^{-1}}$, $2596\pm 42\ \mathrm{km\ s^{-1}}$, $2936\pm 67\ \mathrm{km\ s^{-1}}$, $2911\pm 59\ \mathrm{km\ s^{-1}}$, and $3917\pm 61\ \mathrm{km\ s^{-1}}$, respectively \citep[Table 4 in][]{Feng2024}. Then, the virial product for these lines are $1.24^{+0.16}_{-0.26}\times 10^{7}\ M_{\odot}$, $0.96^{+0.12}_{-0.15}\times 10^{7}\ M_{\odot}$, $1.10^{+0.17}_{-0.24}\times 10^{7}\ M_{\odot}$, $1.05^{+0.16}_{-0.15}\times 10^{7}\ M_{\odot}$, and $0.15^{+0.16}_{-0.15}\times 10^{7}\ M_{\odot}$, respectively. Hence, the virial product of \HeII\ is indeed smaller than that of other lines in NGC 4151 (blue points in Figure~\ref{fig7:mass}). If we add the neglected $\tau_{\mathrm{disk}}$ according to Eq.~\ref{equ:tau_BLR}, the virial product of \HeII\ increases by one order of magnitude and is consistent with that of other BELs (red points in Figure~\ref{fig7:mass}). That is, our results indicate that all BELs have similar $f$. In addition, the SMBH masses of high-redshift AGNs, often estimated by rest-frame UV BEL (e.g., C\,{\sc iv}) single-epoch estimators, may also be affected because these estimators are calibrated to H$\beta$ masses. Hence, understanding the continuum time delays is necessary to improve the precision of SMBH mass measurements. 

\subsection{BLR time delay cosmology} \label{subsec:BLR_cosmology}
The angular size ($\theta$) and linear size ($R_\mathrm{BLR}=c\tau_\mathrm{BLR}$) of BLR provide a new parallax method to measure the angular distances ($D_\mathrm{A}$) of quasars and subsequently measure cosmology model parameters \citep{Elvis2002}. \citet{Wang2020} first measured the angular distance for 3C 273 and constrained the Hubble constant $H_0$, using the angular size measured by spectroastrometry of GRAVITY observations \citep{GRAVITY2018} and the linear size ($c\tau_\mathrm{disk-BLR}$) measured by the BLR RM. Our result suggests that the observed linear size by BLR RM significantly underestimates the actual BLR size (Figure~\ref{fig6:schematic}), which introduces new systematic uncertainties in the angular distance measurement, especially in high-flux states. It is therefore necessary to measure $\tau_\mathrm{disk}$, $\tau_\mathrm{disk-BLR}$, and the BLR angular size simultaneously in the low- and high-flux states, respectively. 

\subsection{Gravitational lensing cosmology} \label{subsec:lense_cosmology}
When a distant quasar is lensed into multiple images by a massive foreground galaxy, the time delay of the flux variations between different images is often used to study cosmological models. The Hubble constant $H_0$ measured by the strong gravitational lensing method has reached a precision of $2.4\%$ \citep[e.g.,][]{Wong2020}. Microlensing caused by stars in the foreground galaxy is one of the sources of uncertainty. The effect of microlensing on the time delay between two images results from the inclination of the accretion disk in the line of sight and the different magnifications of stars in different emission regions of the accretion disk \citep{Tie2018}. The time delay between different emission regions of the accretion disk determines the influence of microlensing on strong-lensing time delays. Our results show that the time delay for NGC 4151 in the high-flux state is $14.9\pm 2.0$ times the SSD model prediction, which can also occur in strong gravitational lensing quasars. The unknown time delay in the emission regions of the accretion disk of strong gravitational lensing quasars introduces an unavoidable uncertainty in gravitational lensing cosmology.

\section{Summary} \label{sec:summary}
We have carried out a 232-day-long $\sim 2$-day-cadence spectroscopic RM campaign on NGC 4151 in the high-flux state. The main conclusions are summarized below. 
\begin{enumerate}
\item We have measured the continuum RM time delays for NGC 4151 in the high-flux state (see Section~\ref{subsubsec:delay_measurement}; Figure~\ref{fig2:light_curve}). Our results are $14.9\pm 2.0$ times larger than the X-ray reprocessed SSD model prediction (Figure~\ref{fig4:tau_evolve}). This is the largest discrepancy detected between the observed time delays and the model predictions. 

\item Our time delays in the high-flux state are $3.8^{+1.8}_{-1.0}$ times larger than those in the low-flux state (see Section~\ref{subsubsec:compare_low}; Figure~\ref{fig4:tau_evolve}). The significant continuum time-delay variations can provide new insight into the multi-wavelength variability of CLAGNs if the time-delay variations are not entirely caused by observing-condition differences (see Section~\ref{subsec:variability}). 

\item Our continuum RM implies that the time delay between optical and ionizing emission varies significantly and can be $\sim 14.9$ times larger than the theoretical expectation. This introduces substantial systematic uncertainties in measuring the BLR linear scales via the BLR RM (see Section~\ref{subsec:BLR_loc}; Figure~\ref{fig6:schematic}). Hence, the continuum time delay will seriously limit the precision of the BLR black-hole mass estimation (see Section~\ref{subsec:BH_mass}), the BLR parallax cosmology (see Section~\ref{subsec:BLR_cosmology}), and gravitational lensing cosmology (see Section~\ref{subsec:lense_cosmology}).
\end{enumerate}

Future simultaneous continuum and BLR RMs can verify our results. Furthermore, theoretical modeling of AGN variability will also provide useful clues to constrain the continuum time delays, thereby improving the BLR size measurements without involving rest-frame UV time-domain observations.

\section*{Acknowledgments}
We thank the anonymous referee for his/her constructive comments that improved the manuscript. We acknowledge S.Y., S.S.W., Y.X.P., Y.J.C., and Y.C.Z. for their helpful discussion. We acknowledge support from the National Key R\&D Program of China (No.~2023YFA1607903, 2021YFA1600404, and No. 2023YFE0101200). S.Y.Z. and M.Y.S. acknowledge support from the National Natural Science Foundation of China (NSFC-12322303), and the Natural Science Foundation of Fujian Province of China (No.~2022J06002). H.C.F. acknowledges support from National Natural Science Foundation of China (NSFC-12203096), Yunnan Fundamental Research Projects (grant NO. 202301AT070339), and Special Research Assistant Funding Project of Chinese Academy of Sciences. S.S.L. acknowledges support from National Natural Science Foundation of China (NSFC-12303022), Yunnan Fundamental Research Projects (grant NO. 202301AT070358), and Yunnan Postdoctoral Research Foundation Funding Project. Y.Q.X. acknowledges support from the National Natural Science Foundation of China (NSFC-12025303 and 12393814). J.X.W. acknowledges support from the National Natural Science Foundation of China (NSFC-12033006 \& 12192221). J.M.B. acknowledges support from National Natural Science Foundation of China (NSFC-11991051), H.T.L. acknowledges support from National Natural Science Foundation of China (NSFC-12373018), K.X.L. acknowledges financial support from the National Natural Science Foundation of China (NSFC-12073068), the Youth Innovation Promotion Association of Chinese Academy of Sciences (2022058), and the Young Talent Project of Yunnan Province. J. M. is supported by the Natural Science Foundation of China (NSFC-12393813), CSST grant, and the Yunnan Revitalization Talent Support Program (Yun-Ling Scholar Project). We acknowledge the support of the staff of the Lijiang 2.4 m telescope. Funding for the telescope has been provided by Chinese Academy of Sciences and the People’s Government of Yunnan Province.

\facilities{YAO:2.4m}

\software{Matplotlib \citep{matplotlib}, Numpy \citep{2020NumPy-Array}, Scipy \citep{2020SciPy-NMeth}, Astropy \citep{Astropy2013}, emcee \citep{emcee-Foreman2013}, PYCCF \citep{Sun2018_pyccf}, statsmodels \citep{seabold2010statsmodels}, astroML \citep{astroML}}

\bibliographystyle{aasjournalv7}
\bibliography{ref.bib}

\end{document}